\documentclass[10pt]{wlscirep}
\usepackage[utf8]{inputenc}
\usepackage[T1]{fontenc}
\usepackage{bm}
\usepackage{newtxmath}
\usepackage{amsmath}
\usepackage{amssymb}
\DeclareMathAlphabet{\mathcal}{OMS}{cmsy}{m}{n}

\title{Imaging arbitrary incoherent source distributions with near quantum-limited resolution}

\author[1]{Erik F. Matlin}
\author[1,*]{Lucas J. Zipp}

\affil[1]{Applied Optics Laboratory, SRI International, Menlo Park, California 94025, USA}

\affil[*]{lucas.zipp@sri.com}

\begin{abstract}
We demonstrate an approach to obtaining near quantum-limited far-field imaging resolution of incoherent sources with arbitrary distributions. Our method assumes no prior knowledge of the source distribution, but rather uses an adaptive approach to imaging via spatial mode demultiplexing that iteratively updates both the form of the spatial imaging modes and the estimate of the source distribution. The optimal imaging modes are determined by minimizing the estimated Cramér-Rao bound over the manifold of all possible sets of orthogonal imaging modes. We have observed through Monte Carlo simulations that the manifold-optimized spatial mode demultiplexing measurement consistently outperforms standard imaging techniques in the accuracy of source reconstructions and comes within a factor of 2 of the absolute quantum limit as set by the quantum Cramér-Rao bound. The adaptive framework presented here allows for a consistent approach to achieving near quantum-limited imaging resolution of arbitrarily distributed sources through spatial mode imaging techniques.
\end{abstract}
\begin{document}

\flushbottom
\maketitle

\thispagestyle{empty}

\section*{Introduction}
Advances in super-resolution imaging methods in the field of fluorescence microscopy \cite{Hell:94,Hell1153,Moerner12596}, as well as techniques that leverage the use of quantum, non-classical light sources \cite{sciarrino_2008,Tenne2019} have demonstrated the ability to surpass the classical diffraction limit. However, these techniques require either active illumination or special sample preparation and are therefore not applicable to passive far-field imaging scenarios. As a result, the resolution limit of standard far-field imaging has remained constrained by the classical diffraction limit, as embodied by the well-known Rayleigh criterion.

Recently, it was shown that the traditional method of far-field imaging, (i.e., detecting photons in the position basis in the image plane) which we will refer to as direct imaging, falls significantly short of the quantum limit of resolution when estimating the separation of two incoherent point sources \cite{Tsang2016}. Instead, the quantum limit is achieved by projecting the light field onto a set of orthogonal spatial modes before detection, sometimes referred to as spatial mode demultiplexing imaging (SPADE). Subsequent work, both theoretical and experimental, has confirmed this approach for achieving quantum-limited localization of two point sources \cite{Nair2016,Lupo2016,Yu2018,Yang2019,Liang2021,Tham2017,Paur2016,Zhou2019,Zhang:20}, including in the presence of noise and crosstalk\cite{Len2020,Oh2021,Gessner2020}, with experimental realizations of SPADE using interferometers \cite{Tham2017}, digital holography\cite{Paur2016,Zhou2019}, and nonlinear techniques\cite{Zhang:20}. In addition, the use of multi-plane light conversion methods has demonstrated the feasibility of demultiplexing a large number of imaging modes \cite{Boucher:20,Fontaine2018,Bade2018}.

Progress has been made in generalizing these techniques to multiple point sources or extended sources \cite{Tsang2017,Bonsma-fisher2019,Bisketzi2019,Dutton,Lupo2020,Pearce2017,PhysRevLett.123.143604}, as well as to adaptive methods \cite{Grace:20,Sajjad_2021,Bao:21}, but no practical framework has been developed for achieving quantum-limited resolution in the case of arbitrary incoherent source distributions, with no prior information about the distribution assumed. This is due to the seemingly daunting challenge of minimizing the Cramér-Rao bound (CRB) for arbitrary sources as the number of parameters tends towards infinity, although progress in certain cases has been made with the use of semiparametric estimation theory \cite{Tsang2019}.

\begin{figure}
\centering
\includegraphics[trim = 0mm 8mm 0mm 0mm,clip,width=13cm]{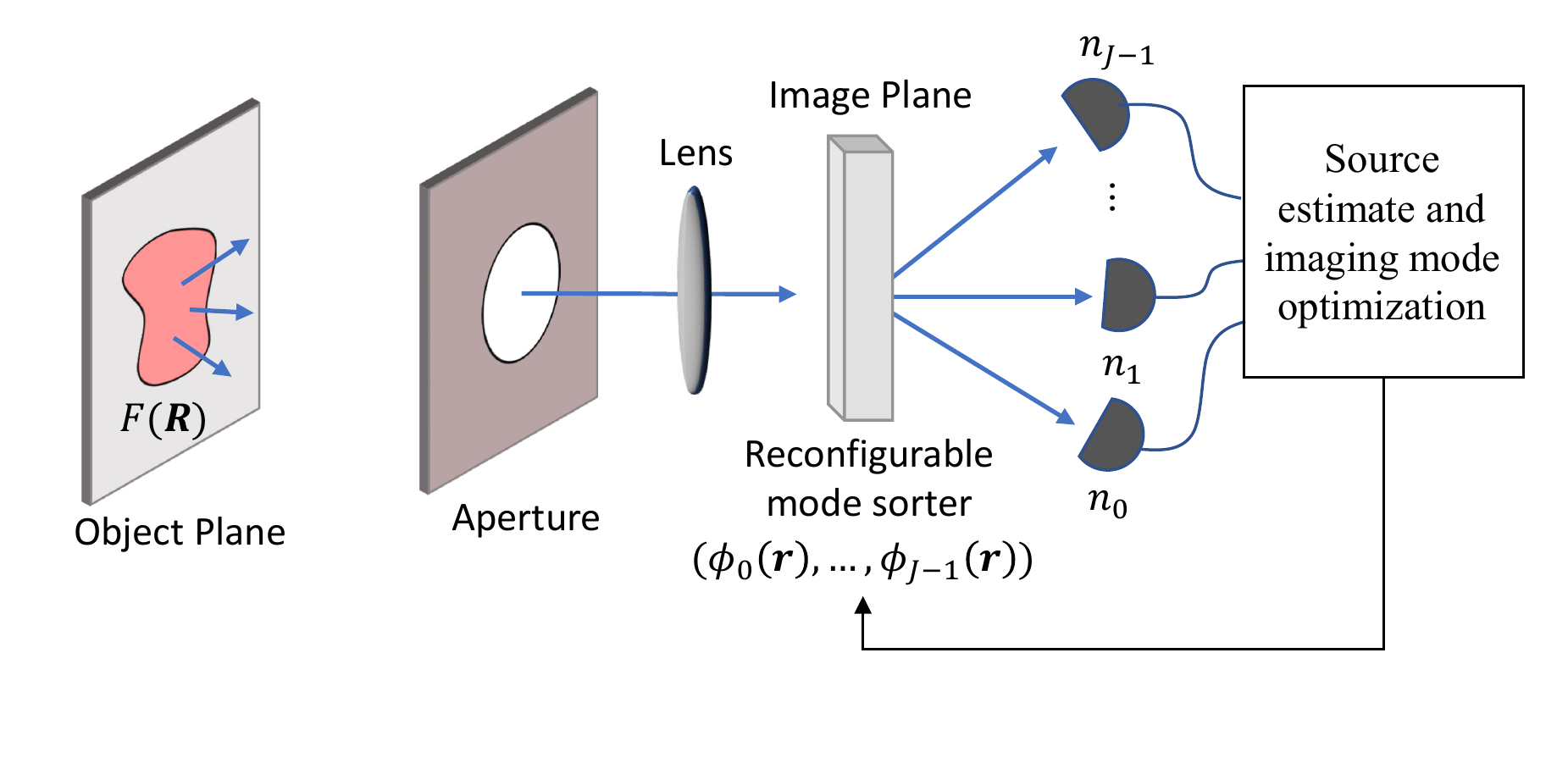}
 \caption{\label{Fig1} Adaptive spatial mode imaging system. Light from the far-field source distribution $F(\bm{R})$ is captured by a hard aperture and imaged onto a spatial mode sorter of imaging modes $\Phi$. At each measurement iteration, the number of photons detected in each channel $n_j$ is used to generate a source estimate, as well as an updated set of optimal spatial imaging modes for the next measurement iteration.}
 \end{figure}

In this work, we demonstrate an adaptive method for achieving near quantum-limited resolution of thermal or incoherent sources based on optimizing the spatial imaging modes over the manifold of orthonormal functions, which we term manifold-optimized spatial mode demultiplexing (MO-SPADE). The method provides an increase in the fundamental resolution of the imaging system, with performance that approaches the quantum limit as set by the quantum Cramér-Rao bound (QCRB). In contrast to super-resolution techniques that apply post-processing to achieve sharper image reconstructions  \cite{Nehme:18,Wang2019}, the measurement method described in this work fundamentally increases the amount of information available on the source distribution.
\paragraph{Optimal imaging methodology.} The setup of the adaptive spatial mode imager, shown in Fig.~\ref{Fig1}, consists of a two-dimensional source region $F(\bm{R})$ emitting light which is imaged in the far-field through a hard aperture. At the image plane, the field is sorted into a set of orthonormal spatial modes $\Phi=\{\phi_j (\bm{r})|j=0,\dots,J-1\}$, and measured by individual photon counting detectors. In this work, we assume a photon source emitting spatially incoherent light with detection events that can be modeled as a Poisson process \cite{Tsang2016}. The photon counts in each channel are then used to reconstruct a source estimate, as well as to update the optimal set of imaging spatial modes for the next measurement iteration. We note that restriction to incoherent sources is not fundamental to the approach, and extension to coherent sources is also possible. 

To optimize the imaging for arbitrary incoherent source distributions, we parametrize the source by decomposing it into a set of orthogonal functions:
\begin{equation}
F(R)=\sum_{k} c_k f_k (R),
\end{equation}
where $F(\bm{R})$ is the source brightness in the object plane, and $f_{k} (\bm{R})$ is the $k_{th}$ function in the source decomposition basis set. We will refer to these functions as source modes, to distinguish them from the spatial imaging modes. To make the problem computationally tractable we restrict the number of source modes to a finite value K, such that $F(\bm{R})$ can be adequately approximated in its subspace. The imaging problem then becomes one of accurately estimating the multiparameter coefficient vector $\bm{c}=(c_1,c_2,\ldots,c_K )$ in the presence of noise. We restrict our attention to the case where the imaging noise is dominated by the photon shot noise and other noise sources are negligible, as is often the case for photon detection in the visible spectrum.
To properly frame this measurement problem, we use the statistical tools of the Fisher information matrix and the corresponding Cramér-Rao bound, which sets the lowest achievable mean squared error (MSE) of the coefficient estimates of any unbiased estimator.
The classical Fisher information per photon measured with the orthonormal imaging modes $\Phi$ is constructed as \cite{Tsang2016}
\begin{equation}\label{FI_eq}
\mathcal{I}(\bm{c};\Phi)_{kl} = \sum_{j=0}^{J-1} \frac{1}{P_j} \left(\frac{\partial P_j}{\partial c_k}\right) \left(\frac{\partial P_j}{\partial c_l}\right) ,
\end{equation}
where $P_j$ is the probability, conditioned on a detection event, of detecting a photon in imaging mode $\phi_j$. For incoherent source distributions this probability is given by
\begin{equation} 
P_j = \int d\bm{R} F(\bm{R}) |\langle \phi_j | \psi_{\bm{R}}^{PSF} \rangle |^2,
\end{equation}
where $|\psi_{\bm{R}}^{PSF} \rangle$ is the field point spread function (PSF) in the image plane originating from source point $\bm{R}$. Replacing $F(\bm{R})$ with its orthogonal decomposition allows us to directly evaluate the partial derivative terms.
The covariance matrix of the coefficient vector for any unbiased estimator satisfies the inequality
\begin{equation} 
\text{cov}(\bm{c}) \geq [\mathcal{I}(\bm{c};\Phi)]^{-1}
\end{equation}
known as the Cramér-Rao bound. Specifically, the MSE of any unbiased estimator of the coefficient $c_k$ is bounded by the corresponding diagonal entry of the Fisher information matrix inverse:
\begin{equation} 
MSE(c_k) \geq[\mathcal{I}(\bm{c};\Phi)]^{-1}_{kk}
\end{equation}
\begin{figure*}
\includegraphics[trim = 0mm 0mm 0mm 0mm, clip,width=\textwidth]{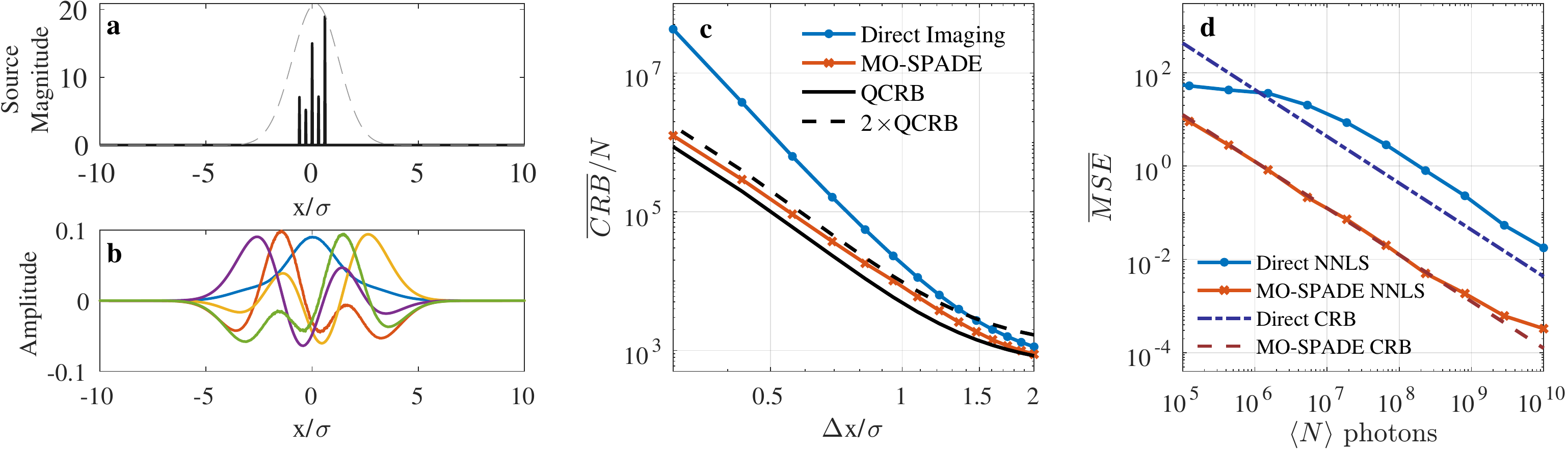}
 \caption{\label{Fig2} Demonstration of MO-SPADE imaging on point sources. (a) Source distribution of 5 equidistant point sources, shown here for a separation of $\Delta x=0.3\sigma$. The dashed curve shows the outline of the PSF-convolved distribution in the image plane. (b) The manifold-optimized spatial imaging modes for the point source distribution.  (c) The CRB averaged over all point source magnitude coefficients plotted at various point source separations. (d) Results of the Monte Carlo simulations for a point source separation of $\Delta x=0.3\sigma$. Each data point represents the MSE of 100 trials averaged over all point source magnitudes.}
 \end{figure*}
The optimal imaging problem can then be cast into one of minimizing the MSE of the source coefficients by minimizing the diagonal elements of the Fisher information matrix inverse. The objective function to be minimized is
\begin{equation} 
L(\Phi)=\mathrm{tr}\left(W[\mathcal{I}(\bm{c};\Phi)]^{-1}\right),
\end{equation}
where $W$ is a weighting matrix that can be applied depending on the nature of the imaging task. In the results shown here, we choose $W$ to be equal to the identity matrix, corresponding to minimizing the mean of the diagonal elements of the Cramér-Rao bound, $\overline{CRB}=1/K\sum_{k=1}^K [\mathcal{I}]^{-1}_{kk} $.
The objective function $L(\Phi)$ is minimized over the manifold of all possible orthonormal imaging modes using numerical manifold optimization methods \cite{Hu2020}. Our specific implementation uses the Scaled Gradient Projection Method (SGPM) designed for the minimization of functions over the set of orthonormal matrices, known as the Stiefel manifold  \cite{Oviedo_2019}.

\section*{Results}
\paragraph{Optimized imaging of point sources and extended sources.}
Before demonstrating the full adaptive imaging method, we first characterize the maximum imaging improvement attainable assuming the true source distribution is known for the purposes of computing the Fisher information matrix in Eq.~(\ref{FI_eq}). We first apply the MO-SPADE imaging method to  one-dimensional imaging of a line of incoherent point sources, where the locations of the sources are known a priori and the task is to estimate the magnitudes of the point sources. This scenario applies, for instance, in estimating the brightness of poorly resolved astronomical objects (e.g., exoplanets), whose locations are known from indirect methods. The source modes are then Dirac delta functions located at the point sources. For simplicity we assume a shift-invariant Gaussian field PSF of $\psi_{R}^{PSF} (x) = (2\pi \sigma^2 )^{-1/4} exp[-(x-R)^2/\left(4\sigma^2\right)]$, where we have normalized the image plane spatial coordinate $x$ by the magnification factor of the imaging system. The width $\sigma$ of the Gaussian PSF is approximately related to the radius $r_c$ of a corresponding Airy disk\cite{Zhang:07} by $\sigma \approx r_{c}/3$.

Figure~\ref{Fig2} displays the results of MO-SPADE imaging of five equally spaced point sources using mode sorting detection with five spatial imaging modes. We find that for point source spacings $\Delta x$ in the sub-Rayleigh regime, MO-SPADE imaging attains a lower CRB compared to direct imaging, with an advantage that grows larger as $\Delta x \rightarrow 0$. In terms of resolution, MO-SPADE allows for smaller $\Delta x$ point source spacings before reaching the equivalent CRB value of direct imaging, showing an increase in effective resolution by over a factor of 1.5 in the deep sub-Rayleigh regime. The spatial imaging modes which attain this performance are shown in Fig.~\ref{Fig2}b. Their profiles are influenced by the source distribution and PSF, and become increasingly complex for larger numbers of imaging modes and sub-Rayleigh source modes.
We confirmed the performance enhancement via Monte Carlo simulations for $\Delta x=0.3 \sigma$, collecting a mean value of $\langle N \rangle$ photons for each trial (Fig.~\ref{Fig2}d). The non-negative least squares (NNLS) estimates of the point source amplitudes follow the CRB limits as expected for large values of  $\langle N \rangle$, and confirms that MO-SPADE imaging achieves an order of magnitude reduction in the MSE of the source reconstruction. For low photon numbers, the MSE of the NNLS estimates falls below the CRB due to its nonzero bias, saturating at a maximum level where the estimation error is of approximately the same magnitude as the source amplitudes.
 
\begin{figure}[t!]
\includegraphics[trim = 0mm 0mm 0mm 0mm, clip,width=\linewidth]{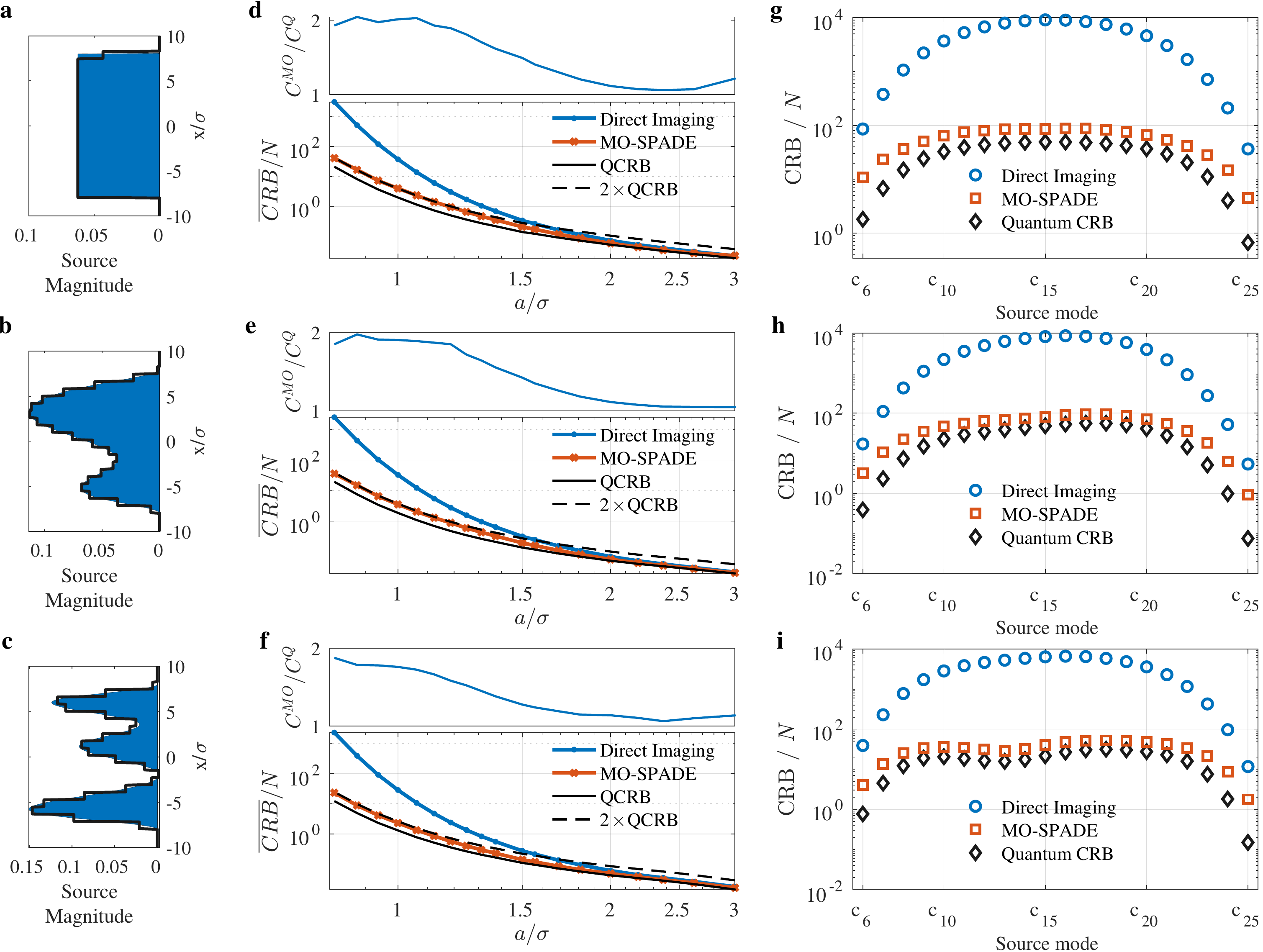}
 \caption{\label{Fig3} Demonstration of MO-SPADE imaging for extended sources. Each row corresponds to analysis of the source distributions shown in (a-c) respectively. The black outline of the sources shows the discretization into rectangle functions with width $a=0.8\sigma$. (d-f) The CRB averaged over all source magnitude coefficients is shown for various source mode rectangle widths $a$. Smaller widths correspond to a larger number of rectangle functions being used to discretize the source region at a finer resolution. The entire discretized source region extends over $\pm12 \sigma$, and $J=K+1$ imaging modes are used. The upper panels plot the ratio between the MO-SPADE CRB ($C^{MO}$) and the quantum CRB ($C^{Q}$). (g-i) CRB of the nonzero source amplitude coefficients ($c_6$ through $c_{25}$) for a discretization width of $a=0.8\sigma$.}
 \end{figure}
 
MO-SPADE imaging also shows enhanced performance when imaging arbitrary extended objects. The source region is approximately discretized into a set of non-overlapping rectangle functions with width $a$. We have chosen rectangle functions for simplicity, but other basis choices may be more appropriate depending on the imaging task. The optimization of the spatial imaging modes $\Phi$ then proceeds in an identical way as with the point source demonstration. Figure~\ref{Fig3} shows the results of MO-SPADE imaging for three different example source distributions of finite extent, including a uniform distribution as well as sources with smoothly varying amplitude variations.
MO-SPADE imaging achieves more than a 10-fold reduction in the CRB of the extended source amplitude estimation in the deep sub-Rayleigh regime, analogous to the point source imaging task. In both of these imaging scenarios, the direct imaging CRB converges towards the optimal CRB for large $a$ or $\Delta x$, in agreement with previous studies on two point sources.
\paragraph{Comparison to the quantum limit.} We also compare our results to the QCRB, which sets a lower bound on the CRB that can be achieved for any possible measurement of a quantum system. We compute the QCRB on an extended source distribution by taking the inverse of the quantum Fisher information matrix $\mathcal{K}_{kl}=\mathrm{Re}\left(\mathrm{tr}\left[\mathcal{L}_{k} (\rho) \mathcal{L}_{l} (\rho)\rho\right]\right)$, where $\mathcal{L}_k (\rho)$ is the symmetric logarithmic derivative (SLD) of density matrix $\rho$ computed as \cite{Tsang2016}:
\begin{equation} 
\mathcal{L}_k(\rho) = \sum_{q,p;\lambda_q+\lambda_p\neq 0} \frac{2}{\lambda_q+\lambda_p} \langle e_q |\frac{\partial \rho}{\partial c_k} | e_p \rangle |e_q\rangle \langle e_p| ,
\end{equation}
where $\lambda_q$ and $|e_q\rangle$ are the eigenvalues and eigenvectors of $\rho$. This version of the quantum Cramér-Rao bound is sometimes referred to as the SLD-CRB, and additional details of the computation are provided in the Supplementary Information.

The multi-parameter QCRB is in general not guaranteed to be attainable by a physically realizable measurement \cite{Liu_2019}. The conditions under which this bound is attainable have been studied in several recent works \cite{Ragy_PhysRevA.94.052108,ALBARELLI2020126311,Sidhu2020,Albarelli2019,Demkowicz_Dobrza_ski_2020,TsangPhysRevX.10.031023}. If the SLD operators commute, that is $[\mathcal{L}_{i},\mathcal{L}_{j}]=0$ for all $i,j$, then the bound is attainable with a projective measurement over the common eigenbasis of the SLD operators. If the SLD operators do not commute, then the CRB of all the parameters cannot be saturated simultaneously with a single projective measurement basis. However, if the system satisfies the condition $\mathrm{tr}(\rho [\mathcal{L}_{i},\mathcal{L}_{j}])=0$, known as weak commutativity, then the quantum bound is still attainable, although collective measurements over multiple copies of the system may be necessary \cite{ALBARELLI2020126311}. In the case of far-field imaging of incoherent sources, the weak commutativity condition holds and the bound is always attainable in principle with the aid of collective measurements \cite{Bisketzi2019}.

Figure~\ref{Fig2}c as well as Fig.~\ref{Fig3}(d-f) show the comparison between the mean CRB of the MO-SPADE imaging and the mean QCRB. We find that in the well-resolved regime the MO-SPADE CRB follows closely the quantum CRB, while in the sub-Rayleigh regime the ratio between the MO-SPADE and quantum CRB approaches, but never exceeds, a factor of 2, as shown in the upper panels of Fig.~\ref{Fig3}(d-f), which plots the ratio between the two CRBs. These results suggest that for far-field incoherent imaging, collective measurements can provide at most a factor of 2 improvement in the CRB compared to the optimal projective measurements as provided by MO-SPADE imaging, even in the deep sub-Rayleigh regime.
\begin{figure}[t]
\centering
\includegraphics[width=\linewidth]{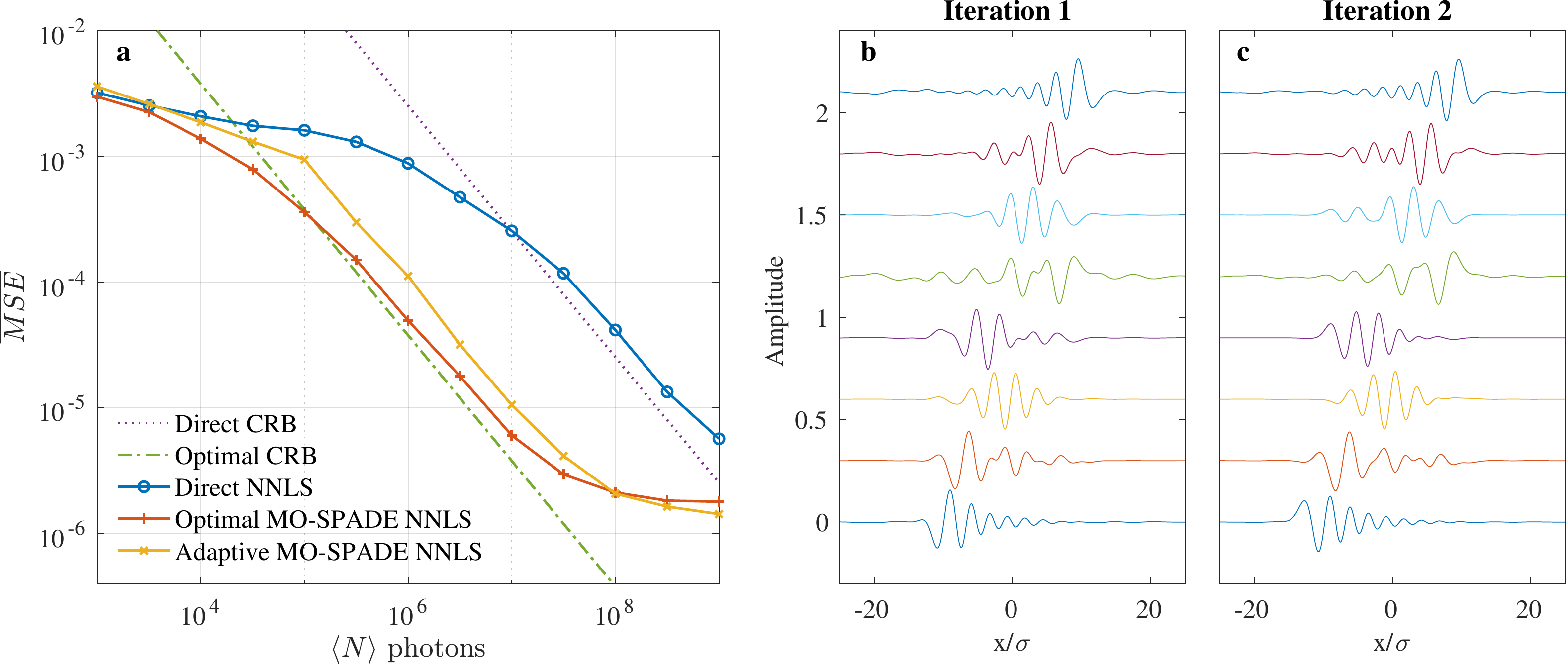}
 \caption{\label{FigMCAdapt} Adaptive imaging simulations. (a) Monte Carlo simulations of adaptive imaging for the source in Fig.~\ref{Fig3}b discretized into widths of $a=0.8\sigma$.  Each data point represents the MSE of 25 trials averaged over all source coefficients. Optimal MO-SPADE refers to the case where prior knowledge of the source is used in optimizing the modes, while adaptive MO-SPADE follows the adaptive imaging procedure outlined in the main text, with no prior knowledge of the source assumed. Only 8 imaging modes are used in the adaptive imaging. Note that at low photon levels, the bias of the NNLS estimator allows the MSE to go below the CRB, while at high photon levels, the MSE of all methods approaches an error floor due to the discretized approximation of the continuous source. (b-c) The imaging modes for the two iterations of the adaptive imaging process for a specific Monte Carlo trial with $\langle N \rangle =10^{6}$ photons.}
 \end{figure}
\paragraph{Adaptive imaging on unknown source distributions.} So far in this work, the CRB and optimal imaging modes have been derived under the assumption of a known source distribution. In most practical cases, the true source distribution is unknown, and it is the purpose of the imaging measurement itself to estimate the source distribution. To solve this problem, an iterative, adaptive approach to imaging is proposed, where the measurement duration is divided into multiple measurement periods, and the source distribution and optimal MO-SPADE imaging modes are alternately estimated. At each adaptive iteration $i$ the estimated Fisher information is calculated using the source estimate $\bm{c_{i-1}^{est}}$ from the previous iteration along with both previous and current imaging modes, weighted according to the relative number of collected photons $N_i$,
\begin{equation} 
\mathcal{I}_{i}^{est} = \sum_{m=1}^{i} N_{m} \mathcal{I}\left(\bm{c_{i-1}^{est}};\Phi_{m}\right) .
\end{equation}
The set of orthonormal imaging modes, $\Phi_i$, are optimized by minimizing the estimated CRB:  $L(\Phi_{i})=\mathrm{tr}\left(W\left[\mathcal{I}_{i}^{est}\right]^{-1}\right)$.
To initialize the adaptive iterations, we set $\Phi_0$ as the direct imaging basis to take the first measurement. At each iteration, all previous measurement data is used in calculating the estimated source coefficients. By following this iterative scheme, we obtain progressively better estimates of the source distribution, and simultaneously approach the optimal imaging modes for the true source distribution. We note that since previous measurement information collected with different imaging bases is included in the source reconstructions, the number of adaptive imaging modes in each iteration can be less than the total number of sources modes. This can potentially allow for large improvements in the imaging while only using a relatively small number of adaptive imaging modes. Additional details on the adaptive algorithm are provided in the Supplementary Information.
\begin{figure}[t]
\centering
\includegraphics[width=\linewidth]{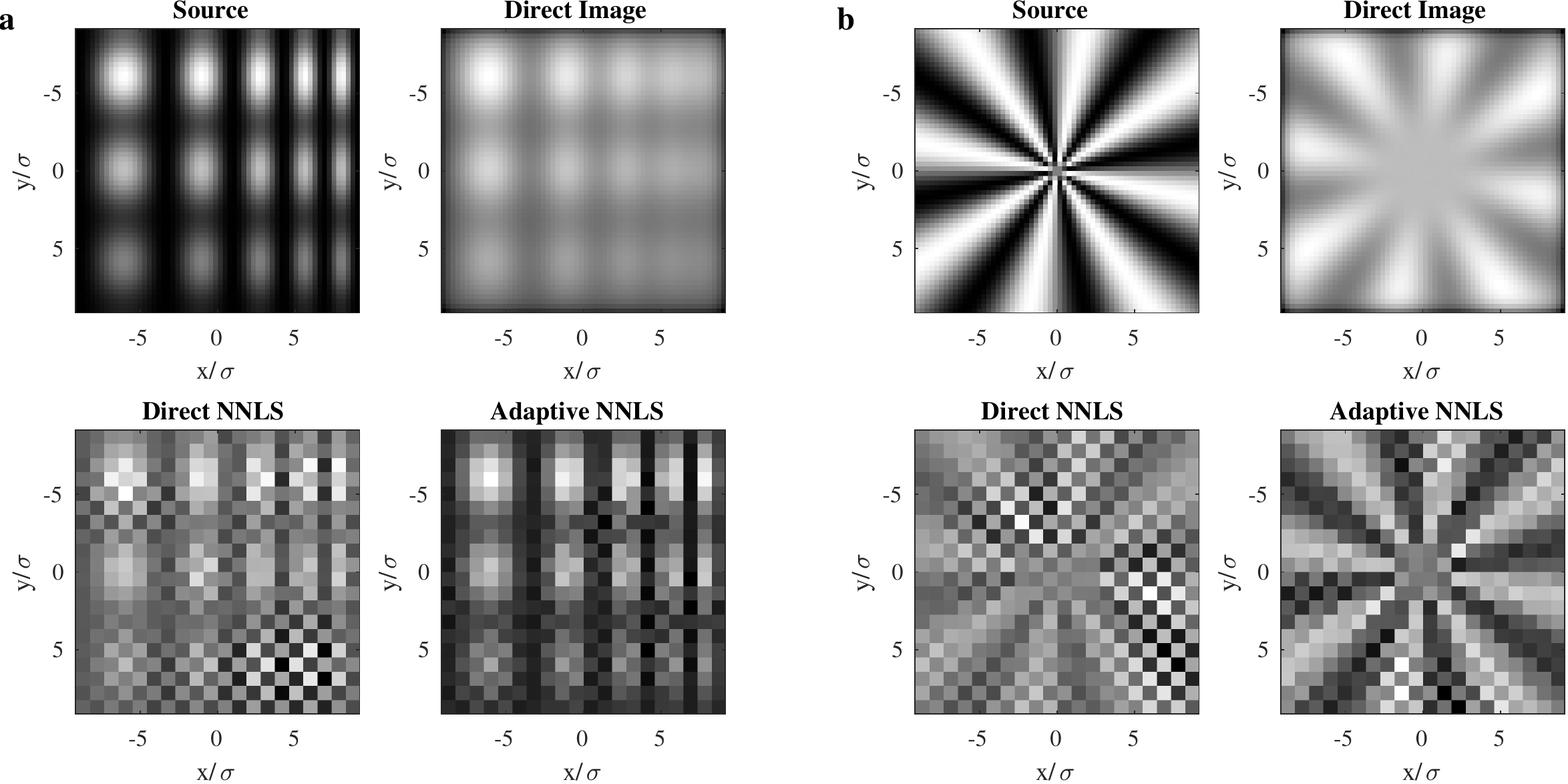}
 \caption{\label{Fig2Dim} Adaptive imaging on low-contrast 2D sources for (a) a chirped source and (b) a Siemens star target. The images correspond to $\langle N \rangle=3\times 10^{13}$ photons collected, and the sources have Michelson contrasts of 0.028 (chirped source) and 0.025 (Siemens star source). For the NNLS reconstructions, the chirped and Siemens star sources are discretized into square bins with a width of $\Delta x \approx 0.9\sigma$. The adaptive reconstruction achieves 4.1 times lower MSE than direct reconstruction for the chirped source, and 7.2 times lower MSE for the Siemens star pattern.}
 \end{figure} 
 
To validate the adaptive approach, we performed a Monte Carlo simulation of the reconstruction performance using the NNLS estimator for the one-dimensional source shown in Fig.~\ref{Fig3}b, discretized into 28 source mode rectangles with width $a=0.8\sigma$. In this simulation, only 8 imaging modes are used for each adaptive step, and the measurement duration is split equally among the initial direct imaging step and two adaptive MO-SPADE measurements -- a configuration that we have empirically found to work well.
The results of this simulation are shown in Fig.~\ref{FigMCAdapt}. At low photon levels, the bias of the NNLS estimator allows the MSE to fall below the respective CRB bounds, and saturates at a maximum error value, as was the case for the point source estimation problem. As the number of photons increases, the MSE begins to follow the CRB, where this trend continues until the approximation of the source by discrete rectangular modes imposes a floor on the reconstruction error. In the CRB-following region, the adaptive MO-SPADE imaging attains a performance that is more than an order of magnitude better than direct imaging and is within a small factor of the optimal (prior-knowledge) MO-SPADE imaging.

An example of the spatial imaging modes which achieve this performance are shown in Fig.~\ref{FigMCAdapt}(b,c). While many of the modes are relatively unchanged between the two iterations, some of the modes exhibit altered oscillation patterns. The small differences between the iterations agrees with our empirical findings that additional iterations of the adaptive imaging procedure beyond the first two tend to produce only marginal improvements in image reconstruction quality.

Finally, we demonstrate the performance of the adaptive imaging on arbitrary two-dimensional source distributions. For these results, we assume a two-dimensional Gaussian PSF with width $\sigma$, corresponding to apodized imaging through a circular aperture. As in the previous one-dimensional example of adaptive imaging, the measurement time is divided evenly among the initial direct imaging step and the two adaptive modal measurements. Figure~\ref{Fig2Dim} shows the results of the adaptive imaging measurement for a source that exhibits spatial frequency chirp and varying contrast, as well as for a Siemens star target. Both sources are placed on top of large uniform backgrounds, simulating the case of low-contrast imaging. The case of low-contrast imaging is particularly advantageous for MO-SPADE imaging, as the bias of the non-negative estimators is lower in the relevant regime where the MSE of the image reconstruction is on the order of the image contrast. Comparing the NNLS reconstruction results, adaptive MO-SPADE imaging outperforms direct imaging and is able to resolve features below the classical Rayleigh resolution limit, with a MSE several times smaller than the direct imaging result. For these plots, the number of imaging modes used in the adaptive modal measurements is equal to the number of source modes plus one, which in both cases equates to 577 imaging modes. However, it is likely that far fewer modes could be used for each adaptive iteration while still obtaining close to optimal results.

\section*{Discussion}
Our results demonstrate a general framework for the imaging of arbitrary distributions with near quantum-limited resolution and accuracy. Further improvements to the adaptive imaging method described here could be made by incorporating the bias of estimators into the optimization framework \cite{Tsang2018}. This would allow for improved imaging performance with estimators that are biased at low photon numbers, such as NNLS and maximum likelihood estimators (e.g. Richardson-Lucy deconvolution). The adaptive imaging method presented here for incoherent thermal sources can be applied to many fields such as multi-emitter fluorescence microscopy and astronomical imaging. Spatial mode imaging is a rapidly developing field, and methods which can incorporate the large number of spatial modes required for adaptive modal imaging of arbitrary 2D sources are being actively investigated \cite{Fontaine2018}. Based on these promising trends, it may be possible to experimentally demonstrate the modal imaging results presented here in the near future.

\section*{Acknowledgements}

We thank Brian Slovick for valuable comments. This work was funded by the Defense Advanced Research Projects Agency (DARPA) under Agreement No. HR00112090126.

\section*{Author contributions}
E.M. and L.Z. developed the concepts, performed the simulations, and wrote the manuscript.

\end{document}